\documentclass[aps,prl,twocolumn,showpacs,superscriptaddress]{revtex4-1}
\usepackage{multirow}
\usepackage[]{graphicx}
\usepackage{amsmath}
\usepackage{amssymb}
\usepackage{amsfonts}
\usepackage{bm}
\usepackage{color}

\begin{document}

\title{Direct measurement of a one-million-dimensional photonic state}

\author{Zhimin Shi}
\email[]{zhiminshi@usf.edu}
\affiliation{Department of Physics, University of South Florida, Tampa, Florida, 33620, USA.}
\author{Mohammad~Mirhosseini}
%\email[]{ideleon@uottawa.ca}
\affiliation{The Institute of Optics, University of Rochester, Rochester, New York 14627 USA}
\author{Jessica Margiewicz}
%\altaffiliation{Y. Wu, J. Margiewicz and Z. Zhu contributed equally to this work.}
\affiliation{Department of Physics, University of South Florida, Tampa, Florida, 33620, USA.}
\author{Mehul~Malik}
\altaffiliation{Institute for Quantum Optics and Quantum Information (IQOQI), Austrian Academy of Sciences, Boltzmanngasse 3, A-1090 Vienna, Austria}
\affiliation{The Institute of Optics, University of Rochester, Rochester, New York 14627 USA}
\author{Freida~Rivera}
\affiliation{Department of Physics, University of South Florida, Tampa, Florida, 33620, USA.}
\author{Robert~W.~Boyd}
\altaffiliation{Department of Physics, University of Ottawa, Ottawa, ON K1N 6N5 Canada}
\affiliation{The Institute of Optics, University of Rochester, Rochester, New York 14627 USA}

\begin{abstract}
Retrieving the vast amount of information carried by a photon is an enduring challenge in quantum metrology science and quantum photonics research. The transverse spatial state of a photon is a convenient high-dimensional quantum system for study, as it has a well-understood classical analogue as the transverse complex field profile of an optical beam. One severe drawback of all currently available quantum metrology techniques is the need for a time-consuming characterization process, which scales very unfavorably with the dimensionality of the quantum system. Here we demonstrate a technique that directly measures a million-dimensional photonic spatial state in a single setting. Through the arrangement of a weak measurement of momentum and parallel strong measurements of position, the complex values of the entire photon state vector become measurable directly. The dimension of our measured state is approximately four orders of magnitude larger than previously measured. Our work opens up a practical route for characterizing high-dimensional quantum systems in real time. Furthermore, our demonstration also serve as a high-speed, extremely-high-resolution unambiguous complex field measurement technique for diverse classical applications.
\end{abstract}

%\ocis{(050.1220)Apertures;(050.1970)Diffractive optics;(100.3190)Inverse problems;(350.6980)Transforms.}

%\pacs{73.20.Mf, 42.65.Hw, 42.70.Nq, 42.65.Re, 42.65.Tg}
\maketitle

Photons play an important role in modern physics as they have a well-understood classical wave picture and a long-perceived particle quanta picture. As a result, photons have been used as a unique quantum platform for studies of quantum science and technology \cite{MandelWolf_CoherentQuantumOpt, Kilin_PiO01:QuantaInfo, Dodonov_JOptB:NonclassicalStateReview}. The transverse wavefunction of a photon \cite{Smith_OL05:Measure_SglPhoton,Lundeen_Nat11:DirectMeasure,Mirhosseini_PRL14:CompressiveDM} is a typical example of a high-dimensional quantum system, which has recently attracted a great amount of research interests for applications in quantum information science including precision measurement\cite{Treps_Science03:QuantumLaserPointer}, high-dimensional entanglement\cite{Mair_Nat01:OAM_entanglement,Langford_PRL04:EntangledQutrits}, parallel information processing \cite{Lassen_PRL07:MultimodeQuanInfo} and secure communication\cite{Mirhosseini_13:SeparationOAM}. For photons in a coherent (pure) state, the transverse wavefunction can be characterized by its state vector, which is a set of complex probability amplitudes expanded over the orthonormal states of a given Hilbert space. The ability to characterize such a high-dimensional quantum state is crucial for fundamental studies of quantum mechanics as well as for manipulating and utilizing single photons for practical applications such as secure communication.

Quantum tomography is an established method used for reconstructing a quantum state through post-processing of the information obtained from a series of strong measurements performed on identically prepared systems \cite{White_PRA01:CharacterizeQuanInfo, Itatani_Nat04:TomographicImaging, Resch_PRL05:ThreePhotonStateTomography, Soderholm_NJP12:PolaTomography, Sych_PRA12:InfoCompleteContVariable, Beck_PRL00:QuanTomoDetArray, Beck_PRL01:QuanMeasureDetArray, Dawes_PRA03:quantumTomowDetArray, Smith_OL05:Measure_SglPhoton}. Recently, direct measurement \cite{Lundeen_Nat11:DirectMeasure} has attracted a tremendous amount of research interest as it offers an alternative metrology technique that can greatly reduces the experimental complexity involved in characterizing a quantum system. The technique of direct measurement has been extended for characterizing various types of quantum systems such as mixed states and high-dimensional states \cite{Lundeen_PRL12:WeakMeasureGeneral, Wu_SciRep:StateTomography, Saivail_NatPhon13:PolarizationDirectMeasure, Malik_13tj:DirectMeasureOAM, Mirhosseini_PRL14:CompressiveDM}.

%it has been shown that one can directly read out the complex probability amplitudes that describe a quantum system through a sequence of weak \cite{Aharonov_PRL88:WeakMeasurementSpin, Duck_PRD89:WeakMeasurement, Ritchie_PRL91:RealizationWeakMeasurement, Johansen_PRL04:WeakMeasurementArbitraryProbeStates, Hosten_Science04:SpinHallEff_weakMeasure, Solli_PRL04:FLSL_GeneralizedWeakValue,DixonPRL09:WeakDeflection,Feizpour_PRL11:WeakSinglePhotonNonlinearity, Kocsis_Science12:EMCCD_entanglement, Dressel_13ti:WeakReview} and strong measurements. Such a procedure, known as ``direct measurement," greatly reduces the experimental complexity involved in characterizing a quantum system, and therefore has attracted a tremendous amount of research interest lately.

%, either in a serial or parallel \cite{} fashion,
To date, all implementations of direct measurement have measured the complex probability amplitudes of a quantum system one at a time. To map out the complete state vector, one would need to perform a sequence of projective measurements at different times, scanning through the bases of the Hilbert space of interest. Hence, the time required to characterize a quantum system scales with the dimension of the system, which makes it difficult to characterize systems of large dimensions. Another drawback of these approaches is the low detection efficiency, because most of the incoming particles are discarded through post-selection during the second step of strong measurement. As a result, the maximum dimension of a quantum state that has been measured using direct measurement is of the order of a hundred \cite{Lundeen_Nat11:DirectMeasure}.

Here we describe a scan-free direct measurement approach that is capable of simultaneously measuring the entire state vector of a pure quantum system, consequently eliminating the need for scanning through each basis state. Specifically, if we wish to measure the state vector in Hilbert space $\mathcal{A}$, we first apply a weak measurement \cite{Aharonov_PRL88:WeakMeasurementSpin, Duck_PRD89:WeakMeasurement, Ritchie_PRL91:RealizationWeakMeasurement, Johansen_PRL04:WeakMeasurementArbitraryProbeStates, Hosten_Science04:SpinHallEff_weakMeasure, Solli_PRL04:FLSL_GeneralizedWeakValue,DixonPRL09:WeakDeflection,Feizpour_PRL11:WeakSinglePhotonNonlinearity, Kocsis_Science12:EMCCD_entanglement, Dressel_13ti:WeakReview} to the quantum system in one fixed state $|b_0\rangle$ of its complementary basis $\mathcal{B}$, and then perform the strong measurement directly in $\mathcal{A}$. Here, a weak measurement refers to the process of applying a weak operator $\hat{\pi}_{a}$ on the system with minimal perturbation such that the original quantum state $|\psi\rangle$ does not collapse fully until a second, conventional (also known as ``strong") measurement is performed. As an example, when we wish to measure the complex probability amplitude of a photon at a certain position $x$, we first perform a weak projection measurement of one particular momentum state ($\hat{\pi}_{p_0}\equiv| p_0 \rangle \langle p_0|$) in $|\psi\rangle$, followed by a strong measurement of the position state $|x\rangle$. Through such a procedure, the measured weak value $\langle \pi_{p} \rangle _{x}^{\rm w}$ is given by (see supplementary material for more details)
\begin{eqnarray}
\langle \pi_{p} \rangle _{x}^{\rm w}  =  \frac{\langle x | p_0 \rangle \langle p_0 | \psi \rangle }{\langle x | \psi \rangle }  =  \frac{e^{-i p_0 x/\hbar} \tilde{\psi} (p_0) }{ \psi(x) },
\end{eqnarray}
where $ \tilde{\psi} (p)$ and $\psi(x)$ denote the state vector of the photon expressed in the momentum and position bases, respectively. When we apply the weak measurement in the zero-momentum state, $p_0=0$, the expression of the weak value simplifies to
\begin{eqnarray}
\langle \pi_{p_0} \rangle_{x}^{\rm w} = \frac{ \nu }{ \psi(x) },
\end{eqnarray}
where $\nu \propto \tilde{\psi} (0)$ is a constant which can be determined through normalizing the state vector.

One sees that the average result of such a measurement directly leads to the complex probability amplitude of the photon at position $x$. The main advantage of our approach is that the weak value $\langle \pi_{p} \rangle _{x}^{\rm w}$ at all positions can be measured simultaneously. This is because the strong measurement in $x$ can be performed on all position states at the same time through the use of an appropriate detector array\cite{Beck_PRL00:QuanTomoDetArray, Beck_PRL01:QuanMeasureDetArray, Dawes_PRA03:quantumTomowDetArray}. Thus, the need for a time-consuming scanning procedure is eliminated, and the entire state vector can be obtained in a single setting.

\begin{figure}[h!]
\centerline{\includegraphics[scale= 1.0]{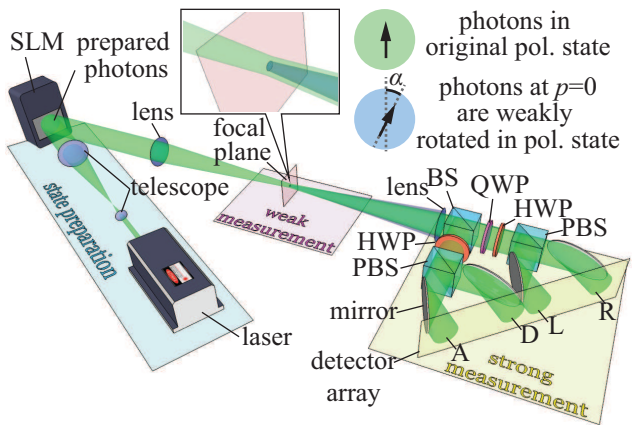}}
\caption{Experimental implementation of a scan-free direct measurement on the transverse spatial state of photons. The photons prepared using a phase-only spatial light modulator (SLM) passes through a 4-$f$ imaging system. The weak measurement is performed in the momentum space, i.e., the common focal plane of the 4-$f$ system, where the linear polarization state of the photons in the zero-momentum state is rotated by a small angle $\alpha$. The strong measurement is performed in the position basis, i.e., the image plane, using a detector array in combination with some polarization optics, where the change in polarization for all position states is measured simultaneously. Specifically, the real and imaginary parts of the weak values are measured in terms of the rotation of the photons' polarization in the diagonal (D)--anti-diagonal (A) linear and left (L)--right (R) handed circular bases respectively, as labeled in the detector array plane. BS: beam splitter; PBS: polarizing beam splitter; HWP: half-wave plate; QWP: quarter-wave plate.}
\label{fig:schematics}
\end{figure}

To demonstrate our scan-free approach, we apply our method to measure the continuous-variable, transverse spatial state of photons. Our experimental procedure is as follows. An ensemble of photons from a collimated laser beam with a fixed polarization state is first prepared using a phase-only spatial light modulator (SLM), which allows a high degree of control of the transverse spatial state $\psi(x)$ of the diffracted photons \cite{Davis_AO99:AmpModwphaseSLM, Arrizon_JOSAA07:SLMgenerateComplexField}. The identically prepared photons pass through a 4$f$ imaging system (see Fig.~1), during which the photons' transverse spatial state expressed in both the momentum and position bases becomes accessible at different locations. First, the weak measurement is performed in the momentum space, i.e., the mutual focal plane of the two lenses, where a second phase-only SLM, in combination with two waveplates, is used to rotate the linear polarization of the photons in the zero-momentum state $| p \rangle =0$ through a small angle $\alpha$. Parallel strong measurements for all the position states are then simultaneously performed at the image plane of the 4$f$ system with a CCD camera, during which the change in the polarization of the photons at each position state is measured. Formally, if we use a two-dimensional vector, $[0 \hspace{0.1cm} 1]^{\rm T}$, to denote the initial polarization state of the photons in the horizontal-vertical linear polarization basis, the complex probability amplitude of the photons in each position state $|x\rangle$ is given by (see supplementary material for detailed derivation)
\begin{eqnarray}
\psi(x) = \langle x | \psi \rangle = \frac{\nu}{\langle \pi_{p} \rangle_{x} ^{\rm w}} = \frac{ \nu'}{ \left[ \langle s_{\rm f}(x) | \hat{\sigma}_1 | s_{\rm f}(x) \rangle - i \langle s_{\rm f}(x) | \hat{\sigma}_2 | s_{\rm f}(x) \rangle \right]},
\label{eq:weakformula}
\end{eqnarray}
where $\hat{\sigma}_1$ and $\hat{\sigma}_2$ are the first and second Pauli operators, respectively, $|s_{\rm f}(x)\rangle$ is the final polarization state at each position $x$ at the image plane, and $\nu$ and $\nu'$ are constants determined through normalization. Note that our specific example of measuring the transverse spatial state of photons can also be fully described using classical language (see supplementary material), as the transverse spatial state of a photon in a pure state is equivalent to the transverse complex field profile of an coherent optical beam in the classical regime. As such, our 4-{\it f} imaging portion of the experimental implementation shares certain similarity with classical point-diffraction interferometry \cite{Smartt_JOSA72:PDI}. However, the quantum mechanical interpretation constitutes the description for a broader range of experiments, and thus can become essential for other quantum systems for which a classical description does not exist.

\begin{figure}[h!]
\centerline{\includegraphics[scale= 1.0]{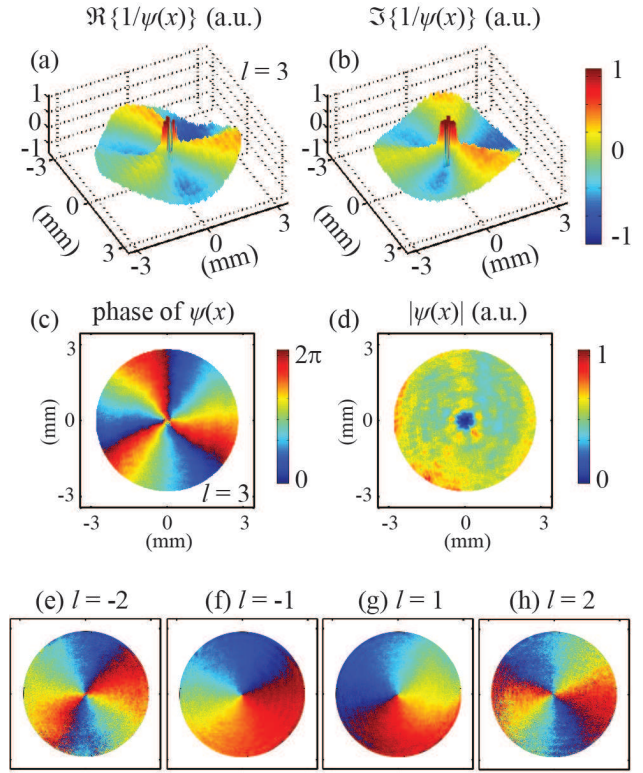}}
\caption{ The measured (a) real and (b) imaginary parts of the two-dimensional weak values and the corresponding (c) phase and (d) amplitude profiles of photons carrying orbital angular momentum (OAM) with quantum number $l=3$. The measured weak values have very large magnitude towards the center of the mode and therefore is truncated for better visualization purposes. (e)-(h): The extracted phase profile of photons carrying OAM with quantum number $l$ ranging from -2 to 2.}
\label{fig:OAM_data}
\end{figure}

In our experimental demonstration, we first characterize photons carrying orbital angular momentum (OAM)\cite{Yao_AOP11:OAMreview}, which has recently been the subject of many fundamental studies in quantum mechanics \cite{Mair_Nat01:OAM_entanglement,Leach_Science10:QuantumCorrelationOAM,Dada_NatPhys11:HighDimEntanglement,Malik_OE12:Turbulence,Mirhosseini_13:SeparationOAM}. We generate photons carrying different values of OAM quantum number $l$ using the SLM technique described above. The real and imaginary parts of the measured weak value $\langle \pi_{p} \rangle_{x} ^{\rm w}$ for photons with $l=3$ are plotted in Fig.~\ref{fig:OAM_data}(a) and (b), respectively. One sees that the magnitude of $|\langle \pi_{p} \rangle_{x} ^{\rm w}|$ becomes very large towards the center of the OAM beam, which is exactly expected due to the inverse relation between $\langle \pi_{p} \rangle_{x} ^{\rm w}$ and the complex probability amplitude $\phi(x)$ [cf. Eq.~(\ref{eq:weakformula})] of an OAM beam, which approaches zero towards the phase singularity at the center. The corresponding phase and amplitude of $|\psi(x)|$ is shown in Fig.~\ref{fig:OAM_data}(c) and (d), which accurately reveals the azimuthal phase structure and the central-null feature of the amplitude. We further quantify the fidelity of our measurement result using the well-defined ``transition probability'' $F \equiv | \langle \pi_{\rm exp} | \pi_{\rm ide} \rangle |^2$ where $|\pi_{\rm exp}\rangle$ and $| \pi_{\rm ide} \rangle$ denotes the experimentally measured and the ideal photon states, respectively. The fidelity of the shown $l=3$ OAM mode in the spatial Hilbert space is calculated to be approximately 0.86. Note that the nonideal optical system we use to generate the photon state also contribute partially to the non-unity value of our measured fidelity. Nonetheless, the high fidelity of our result demonstrates that our direct measurement technique is indeed capable of measuring the complex-value quantum state vector with very high accuracy. Similarly high-quality results are obtained for photons carrying other quantum numbers of OAM, and the measured phase profile of the OAM modes with $l$ ranging from $-2$ to $2$ is shown in Fig.~\ref{fig:OAM_data}(e)-(h). Note that the OAM modes does not constitute the Hilbert space of study here, but are used rather as examples of arbitrary transverse spatial state the photons can be in. We have the full control of the complex probability amplitude of the photon at each spatial points  The dimensionality of our measured state is approximately 1.2 million, which is determined by the spatial extent of the photons (approximately 7 mm in diameter) and the discretization of our detector array (with pixel size of 5.4 $\mu$m$^2$). The effective dimensionality of the measured continuous-basis position space is reduced to a fraction of a million due to the space-bandwidth-product of our imaging system. Yet, the effective dimensionality of the measured Hilbert space can be arbitrarily enlarged by optimizing the measurement apparatus, such as using larger optical components and a larger-area detector array.

\begin{figure}[h!]
\centerline{\includegraphics[scale= 1.0]{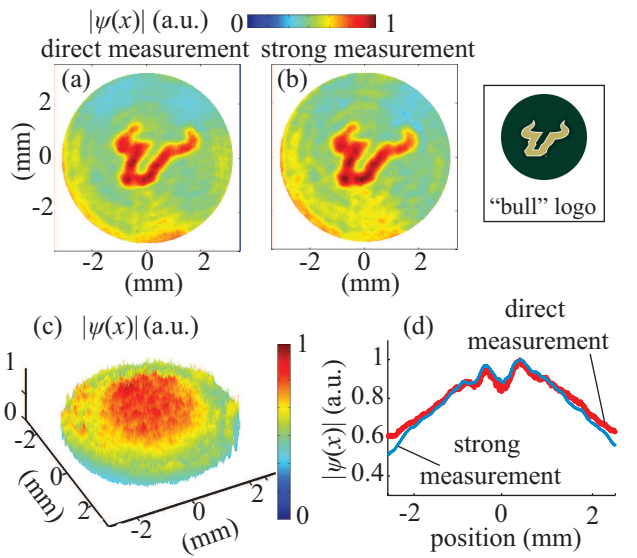}}
\caption{Upper row: The measured magnitude of the probability amplitude $|\psi(x)|$ of photons with an amplitude profile incorporating a University of South Florida ``Bull'' logo using (a) our direct measurement approach and (b) conventional strong measurement. Lower row: (c) The measured magnitude of the probability amplitude $|\psi(x)|$ of photons with a truncated Gaussian amplitude profile using our scan-free direct measurement approach; and (d) one cross-section of the directly measured result (thick red line) in comparison with result of the conventional strong (intensity) measurement (thin blue line). The actual $|\psi(x)|$ profile is the result of diffraction and propagation of the photons through our non-ideal imaging system.}
\label{fig:Amplitude}
\end{figure}

We then test our method on photons with more arbitrary transverse state profiles. First, we impose a bull-shaped letter ``U'' pattern on the amplitude profile of the photons and with various Zernike phase profiles. The obtained magnitude of the probability amplitude $|\psi(x)|$ is shown in Fig.~\ref{fig:Amplitude}(a), which is in good agreement with the result obtained using conventional intensity (strong) measurements [see Fig.~\ref{fig:Amplitude}(b)], i.e., the square root of a direct image captured by the camera. We also measure photons with a gradually-varying amplitude profile carrying various Zernike polynomial phase structures. One measured $|\psi(x)|$ using our direct approach is shown in Fig.~\ref{fig:Amplitude}(c), and a cross section of $|\psi(x)|$ (the thick red line) is plotted in Fig.~\ref{fig:Amplitude}(d) in comparison with the conventional strong measurement result (the thin blue line). Note that the theory of our approach assumes that the perturbation due to the weak measurement in the momentum space is sufficiently weak that the rotation of the polarization state of photons in each position state is small. This imposes a practical limit on the minimum probability ($|\psi(x)|^2$) that can be accurately measured, which is experimentally determined by the accuracy of the polarization measurement in our case.

Since the expectation values of $\langle s_{\rm f}(x) | \hat{\sigma}_1 | s_{\rm f}(x) \rangle$ and $\langle s_{\rm f}(x) | \hat{\sigma}_2 | s_{\rm f}(x) \rangle$ at all position states are measured in parallel, our approach is capable of monitoring the dynamic variation of the complex amplitude profile of an ensemble of photons, either in coherent states or in single-photon states \cite{Fickler_SciRep:RealTimeImagingEntanglement}, in real time. To illustrate such capability, we impose a dynamically-changing phase profile on the photons with a constant amplitude within a circular aperture. The encoded phase structure switches among various rotating Zernike polynomial functions, and the dynamic evolution of the complex probability of the photons is recorded continuously using the camera in movie-shooting mode at 14 frames per second. The measured dynamical variation of the phase profile of the photons is shown in the supplementary movie S1, which accurately reveals the designed variation pattern.

One should note that even though we measure a large ensemble of identically-prepared photons in our experiment, our procedure determines the complex transverse spatial state of each photon, as has been demonstrated in previous direct measurement studies \cite{Lundeen_Nat11:DirectMeasure, Saivail_NatPhon13:PolarizationDirectMeasure}. Meanwhile, both the approach outlined here and the experimental apparatus are directly applicable to identically-prepared single photons, provided that we use detector arrays that are capable of detecting single photons with high quantum efficiency such as SPAD arrays, commercial cooled CCD cameras\cite{Dawes_PRA03:quantumTomowDetArray, Kocsis_Science12:EMCCD_entanglement}, electron-multiplying CCD cameras \cite{Edgar_NatComm12:EMCCD_entanglement} or intensified CCD cameras \cite{Fickler_SciRep:RealTimeImagingEntanglement}. Note that lower quantum efficiency of single-photon detectors would require summing over larger number of measurements on identically-prepared single photons to average out the noise, but the complex-valued state-vector of single photons can still be measured directly in a single setting through the same procedure.

The number of incoming photons needed to map out an entire state vector of dimension $N$ using our scan-free approach is comparable to the number of photons needed to measure the complex probability amplitude of at a single position using the previous direct measurement approach \cite{Lundeen_Nat11:DirectMeasure}. This can also be understood by the fact that most of the incoming photons are discarded through post-selection in previous direct measurement approaches whereas our scan-free approach does not involve any post selection. Thus, for a quantum system with a dimensionality of $N$, our approach is approximately $N$ times more efficient as compared to a state-by-state scanning approach (see supplementary material for details).

Our scan-free direct measurement approach can be extended to measure the state of other quantum systems in a straightforward fashion, and it opens up the possibility to characterizing a high-dimensional quantum system in real-time for which a state-by-state scanning process would become impractically time-consuming or even infeasible. Moreover, our specific demonstration of measuring photons' transverse spatial state can be readily used to measure directly the phase profile of an optical beam, and therefore is also a promising new technology for classical wavefront sensing applications in fields as diverse as observational astronomy, free-space optical communication, and biomedical imaging.

\section*{Funding Information}
Mo.M., Me.M. and R.W.B. acknowledge the support from the DARPA InPho Program.  In addition, R.W.B. acknowledges support from the Canada Excellence Research Chairs program, and Me.M. acknowledges support from the European Commission through a Marie Curie fellowship.

\section*{Acknowledgments}
The authors thank M. Lavery and J. Leach for helpful discussions.

\section*{Supplemental Documents}

%\bigskip \noindent See \href{link}{Supplement 1} for supporting content.

\bigskip \noindent See Supplement 1 and Supplement movie S1 for supporting content.

%
%\emph{Optica} authors may include supplemental documents with the primary manuscript. For details, see \href{http://www.opticsinfobase.org/submit/style/supplementary-materials-optica.cfm}{Supplementary Materials in Optica}. To reference the supplementary document, the statement ``See Supplement 1 for supporting content.'' should appear at the bottom of the manuscript (above the references).

%\bigskip \noindent See \href{link}{Supplement 1} for supporting content.

%\noindent Add citations manually or use BibTeX. See \cite{Zhang:14}.

% Bibliography
%\bibliography{Bib_PRINT_quan,Bib_wavefrontSensing} %\bibliography{sample}

\begin{thebibliography}{42}%
\makeatletter
\providecommand \@ifxundefined [1]{%
 \@ifx{#1\undefined}
}%
\providecommand \@ifnum [1]{%
 \ifnum #1\expandafter \@firstoftwo
 \else \expandafter \@secondoftwo
 \fi
}%
\providecommand \@ifx [1]{%
 \ifx #1\expandafter \@firstoftwo
 \else \expandafter \@secondoftwo
 \fi
}%
\providecommand \natexlab [1]{#1}%
\providecommand \enquote  [1]{``#1''}%
\providecommand \bibnamefont  [1]{#1}%
\providecommand \bibfnamefont [1]{#1}%
\providecommand \citenamefont [1]{#1}%
\providecommand \href@noop [0]{\@secondoftwo}%
\providecommand \href [0]{\begingroup \@sanitize@url \@href}%
\providecommand \@href[1]{\@@startlink{#1}\@@href}%
\providecommand \@@href[1]{\endgroup#1\@@endlink}%
\providecommand \@sanitize@url [0]{\catcode `\\12\catcode `\$12\catcode
  `\&12\catcode `\#12\catcode `\^12\catcode `\_12\catcode `\%12\relax}%
\providecommand \@@startlink[1]{}%
\providecommand \@@endlink[0]{}%
\providecommand \url  [0]{\begingroup\@sanitize@url \@url }%
\providecommand \@url [1]{\endgroup\@href {#1}{\urlprefix }}%
\providecommand \urlprefix  [0]{URL }%
\providecommand \Eprint [0]{\href }%
\providecommand \doibase [0]{http://dx.doi.org/}%
\providecommand \selectlanguage [0]{\@gobble}%
\providecommand \bibinfo  [0]{\@secondoftwo}%
\providecommand \bibfield  [0]{\@secondoftwo}%
\providecommand \translation [1]{[#1]}%
\providecommand \BibitemOpen [0]{}%
\providecommand \bibitemStop [0]{}%
\providecommand \bibitemNoStop [0]{.\EOS\space}%
\providecommand \EOS [0]{\spacefactor3000\relax}%
\providecommand \BibitemShut  [1]{\csname bibitem#1\endcsname}%
\let\auto@bib@innerbib\@empty
%</preamble>
\bibitem [{\citenamefont {Mandel}\ and\ \citenamefont
  {Wolf}(1995)}]{MandelWolf_CoherentQuantumOpt}%
  \BibitemOpen
  \bibfield  {author} {\bibinfo {author} {\bibfnamefont {L.}~\bibnamefont
  {Mandel}}\ and\ \bibinfo {author} {\bibfnamefont {E.}~\bibnamefont {Wolf}},\
  }\enquote {\bibinfo {title} {Optical coherence and quantum optics},}\ \
  (\bibinfo  {publisher} {Plenum Press},\ \bibinfo {address} {New York},\
  \bibinfo {year} {1995})\ \bibinfo {edition} {1st}\ ed.\BibitemShut {Stop}%
\bibitem [{\citenamefont {Kilin}(2001)}]{Kilin_PiO01:QuantaInfo}%
  \BibitemOpen
  \bibfield  {author} {\bibinfo {author} {\bibfnamefont {S.}~\bibnamefont
  {Kilin}},\ }in\ \href@noop {} {\emph {\bibinfo {booktitle} {Progress in
  Optics}}},\ Vol.~\bibinfo {volume} {42},\ \bibinfo {editor} {edited by\
  \bibinfo {editor} {\bibfnamefont {E.}~\bibnamefont {Wolf}}}\ (\bibinfo
  {publisher} {Elsevier Science},\ \bibinfo {address} {Amsterdam},\ \bibinfo
  {year} {2001})\ pp.\ \bibinfo {pages} {1--90}\BibitemShut {NoStop}%
\bibitem [{\citenamefont
  {Dodonov}(2002)}]{Dodonov_JOptB:NonclassicalStateReview}%
  \BibitemOpen
  \bibfield  {author} {\bibinfo {author} {\bibfnamefont {V.~V.}\ \bibnamefont
  {Dodonov}},\ }\href {http://stacks.iop.org/1464-4266/4/i=1/a=201} {\bibfield
  {journal} {\bibinfo  {journal} {J. Opt. B: Quantum and Semiclassical Opt.}\
  }\textbf {\bibinfo {volume} {4}},\ \bibinfo {pages} {R1} (\bibinfo {year}
  {2002})}\BibitemShut {NoStop}%
\bibitem [{\citenamefont {Smith}\ \emph {et~al.}(2005)\citenamefont {Smith},
  \citenamefont {Killett}, \citenamefont {Raymer}, \citenamefont {Walmsley},\
  and\ \citenamefont {Banaszek}}]{Smith_OL05:Measure_SglPhoton}%
  \BibitemOpen
  \bibfield  {author} {\bibinfo {author} {\bibfnamefont {B.~J.}\ \bibnamefont
  {Smith}}, \bibinfo {author} {\bibfnamefont {B.}~\bibnamefont {Killett}},
  \bibinfo {author} {\bibfnamefont {M.~G.}\ \bibnamefont {Raymer}}, \bibinfo
  {author} {\bibfnamefont {I.~A.}\ \bibnamefont {Walmsley}}, \ and\ \bibinfo
  {author} {\bibfnamefont {K.}~\bibnamefont {Banaszek}},\ }\href {\doibase
  10.1364/OL.30.003365} {\bibfield  {journal} {\bibinfo  {journal} {Opt.
  Lett.}\ }\textbf {\bibinfo {volume} {30}},\ \bibinfo {pages} {3365} (\bibinfo
  {year} {2005})}\BibitemShut {NoStop}%
\bibitem [{\citenamefont {Lundeen}\ \emph {et~al.}(2011)\citenamefont
  {Lundeen}, \citenamefont {Sutherland}, \citenamefont {Patel}, \citenamefont
  {Stewart},\ and\ \citenamefont {Bamber}}]{Lundeen_Nat11:DirectMeasure}%
  \BibitemOpen
  \bibfield  {author} {\bibinfo {author} {\bibfnamefont {J.~S.}\ \bibnamefont
  {Lundeen}}, \bibinfo {author} {\bibfnamefont {B.}~\bibnamefont {Sutherland}},
  \bibinfo {author} {\bibfnamefont {A.}~\bibnamefont {Patel}}, \bibinfo
  {author} {\bibfnamefont {C.}~\bibnamefont {Stewart}}, \ and\ \bibinfo
  {author} {\bibfnamefont {C.}~\bibnamefont {Bamber}},\ }\href@noop {}
  {\bibfield  {journal} {\bibinfo  {journal} {Nature}\ }\textbf {\bibinfo
  {volume} {474}},\ \bibinfo {pages} {188} (\bibinfo {year}
  {2011})}\BibitemShut {NoStop}%
\bibitem [{\citenamefont {Mirhosseini}\ \emph {et~al.}(2014)\citenamefont
  {Mirhosseini}, \citenamefont {Maga\~na Loaiza}, \citenamefont
  {Hashemi~Rafsanjani},\ and\ \citenamefont
  {Boyd}}]{Mirhosseini_PRL14:CompressiveDM}%
  \BibitemOpen
  \bibfield  {author} {\bibinfo {author} {\bibfnamefont {M.}~\bibnamefont
  {Mirhosseini}}, \bibinfo {author} {\bibfnamefont {O.~S.}\ \bibnamefont
  {Maga\~na Loaiza}}, \bibinfo {author} {\bibfnamefont {S.~M.}\ \bibnamefont
  {Hashemi~Rafsanjani}}, \ and\ \bibinfo {author} {\bibfnamefont {R.~W.}\
  \bibnamefont {Boyd}},\ }\href {\doibase 10.1103/PhysRevLett.113.090402}
  {\bibfield  {journal} {\bibinfo  {journal} {Phys. Rev. Lett.}\ }\textbf
  {\bibinfo {volume} {113}},\ \bibinfo {pages} {090402} (\bibinfo {year}
  {2014})}\BibitemShut {NoStop}%
\bibitem [{\citenamefont {Treps}\ \emph {et~al.}(2003)\citenamefont {Treps},
  \citenamefont {Grosse}, \citenamefont {Bowen}, \citenamefont {Fabre},
  \citenamefont {Bachor},\ and\ \citenamefont
  {Lam}}]{Treps_Science03:QuantumLaserPointer}%
  \BibitemOpen
  \bibfield  {author} {\bibinfo {author} {\bibfnamefont {N.}~\bibnamefont
  {Treps}}, \bibinfo {author} {\bibfnamefont {N.}~\bibnamefont {Grosse}},
  \bibinfo {author} {\bibfnamefont {W.~P.}\ \bibnamefont {Bowen}}, \bibinfo
  {author} {\bibfnamefont {C.}~\bibnamefont {Fabre}}, \bibinfo {author}
  {\bibfnamefont {H.-A.}\ \bibnamefont {Bachor}}, \ and\ \bibinfo {author}
  {\bibfnamefont {P.~K.}\ \bibnamefont {Lam}},\ }\href@noop {} {\bibfield
  {journal} {\bibinfo  {journal} {Science}\ }\textbf {\bibinfo {volume}
  {301}},\ \bibinfo {pages} {940} (\bibinfo {year} {2003})}\BibitemShut
  {NoStop}%
\bibitem [{\citenamefont {Mair}\ \emph {et~al.}(2001)\citenamefont {Mair},
  \citenamefont {Vaziri}, \citenamefont {Weihs},\ and\ \citenamefont
  {Zeilinger}}]{Mair_Nat01:OAM_entanglement}%
  \BibitemOpen
  \bibfield  {author} {\bibinfo {author} {\bibfnamefont {A.}~\bibnamefont
  {Mair}}, \bibinfo {author} {\bibfnamefont {A.}~\bibnamefont {Vaziri}},
  \bibinfo {author} {\bibfnamefont {G.}~\bibnamefont {Weihs}}, \ and\ \bibinfo
  {author} {\bibfnamefont {A.}~\bibnamefont {Zeilinger}},\ }\href@noop {}
  {\bibfield  {journal} {\bibinfo  {journal} {Nature}\ }\textbf {\bibinfo
  {volume} {412}},\ \bibinfo {pages} {313} (\bibinfo {year}
  {2001})}\BibitemShut {NoStop}%
\bibitem [{\citenamefont {Langford}\ \emph {et~al.}(2004)\citenamefont
  {Langford}, \citenamefont {Dalton}, \citenamefont {Harvey}, \citenamefont
  {O'Brien}, \citenamefont {Pryde}, \citenamefont {Gilchrist}, \citenamefont
  {Bartlett},\ and\ \citenamefont {White}}]{Langford_PRL04:EntangledQutrits}%
  \BibitemOpen
  \bibfield  {author} {\bibinfo {author} {\bibfnamefont {N.~K.}\ \bibnamefont
  {Langford}}, \bibinfo {author} {\bibfnamefont {R.~B.}\ \bibnamefont
  {Dalton}}, \bibinfo {author} {\bibfnamefont {M.~D.}\ \bibnamefont {Harvey}},
  \bibinfo {author} {\bibfnamefont {J.~L.}\ \bibnamefont {O'Brien}}, \bibinfo
  {author} {\bibfnamefont {G.~J.}\ \bibnamefont {Pryde}}, \bibinfo {author}
  {\bibfnamefont {A.}~\bibnamefont {Gilchrist}}, \bibinfo {author}
  {\bibfnamefont {S.~D.}\ \bibnamefont {Bartlett}}, \ and\ \bibinfo {author}
  {\bibfnamefont {A.~G.}\ \bibnamefont {White}},\ }\href {\doibase
  10.1103/PhysRevLett.93.053601} {\bibfield  {journal} {\bibinfo  {journal}
  {Phys. Rev. Lett.}\ }\textbf {\bibinfo {volume} {93}},\ \bibinfo {pages}
  {053601} (\bibinfo {year} {2004})}\BibitemShut {NoStop}%
\bibitem [{\citenamefont {Lassen}\ \emph {et~al.}(2007)\citenamefont {Lassen},
  \citenamefont {Delaubert}, \citenamefont {Janousek}, \citenamefont {Wagner},
  \citenamefont {Bachor}, \citenamefont {Lam}, \citenamefont {Treps},
  \citenamefont {Buchhave}, \citenamefont {Fabre},\ and\ \citenamefont
  {Harb}}]{Lassen_PRL07:MultimodeQuanInfo}%
  \BibitemOpen
  \bibfield  {author} {\bibinfo {author} {\bibfnamefont {M.}~\bibnamefont
  {Lassen}}, \bibinfo {author} {\bibfnamefont {V.}~\bibnamefont {Delaubert}},
  \bibinfo {author} {\bibfnamefont {J.}~\bibnamefont {Janousek}}, \bibinfo
  {author} {\bibfnamefont {K.}~\bibnamefont {Wagner}}, \bibinfo {author}
  {\bibfnamefont {H.-A.}\ \bibnamefont {Bachor}}, \bibinfo {author}
  {\bibfnamefont {P.~K.}\ \bibnamefont {Lam}}, \bibinfo {author} {\bibfnamefont
  {N.}~\bibnamefont {Treps}}, \bibinfo {author} {\bibfnamefont
  {P.}~\bibnamefont {Buchhave}}, \bibinfo {author} {\bibfnamefont
  {C.}~\bibnamefont {Fabre}}, \ and\ \bibinfo {author} {\bibfnamefont {C.~C.}\
  \bibnamefont {Harb}},\ }\href {\doibase 10.1103/PhysRevLett.98.083602}
  {\bibfield  {journal} {\bibinfo  {journal} {Phys. Rev. Lett.}\ }\textbf
  {\bibinfo {volume} {98}},\ \bibinfo {pages} {083602} (\bibinfo {year}
  {2007})}\BibitemShut {NoStop}%
\bibitem [{\citenamefont {Mirhosseini}\ \emph {et~al.}(2013)\citenamefont
  {Mirhosseini}, \citenamefont {Malik}, \citenamefont {Shi},\ and\
  \citenamefont {Boyd}}]{Mirhosseini_13:SeparationOAM}%
  \BibitemOpen
  \bibfield  {author} {\bibinfo {author} {\bibfnamefont {M.}~\bibnamefont
  {Mirhosseini}}, \bibinfo {author} {\bibfnamefont {M.}~\bibnamefont {Malik}},
  \bibinfo {author} {\bibfnamefont {Z.}~\bibnamefont {Shi}}, \ and\ \bibinfo
  {author} {\bibfnamefont {R.~W.}\ \bibnamefont {Boyd}},\ }\href@noop {}
  {\bibfield  {journal} {\bibinfo  {journal} {Nature Communications}\ }\textbf
  {\bibinfo {volume} {4}},\ \bibinfo {pages} {2781} (\bibinfo {year}
  {2013})}\BibitemShut {NoStop}%
\bibitem [{\citenamefont {White}\ \emph {et~al.}(2001)\citenamefont {White},
  \citenamefont {James}, \citenamefont {Munro},\ and\ \citenamefont
  {Kwiat}}]{White_PRA01:CharacterizeQuanInfo}%
  \BibitemOpen
  \bibfield  {author} {\bibinfo {author} {\bibfnamefont {A.~G.}\ \bibnamefont
  {White}}, \bibinfo {author} {\bibfnamefont {D.~F.~V.}\ \bibnamefont {James}},
  \bibinfo {author} {\bibfnamefont {W.~J.}\ \bibnamefont {Munro}}, \ and\
  \bibinfo {author} {\bibfnamefont {P.~G.}\ \bibnamefont {Kwiat}},\ }\href
  {\doibase 10.1103/PhysRevA.65.012301} {\bibfield  {journal} {\bibinfo
  {journal} {Phys. Rev. A}\ }\textbf {\bibinfo {volume} {65}},\ \bibinfo
  {pages} {012301} (\bibinfo {year} {2001})}\BibitemShut {NoStop}%
\bibitem [{\citenamefont {Itatani}\ \emph {et~al.}(2004)\citenamefont
  {Itatani}, \citenamefont {Levesque}, \citenamefont {Zeidler}, \citenamefont
  {Niikura}, \citenamefont {P\'{e}pin}, \citenamefont {Kieffer}, \citenamefont
  {Corkum},\ and\ \citenamefont
  {Villeneuve}}]{Itatani_Nat04:TomographicImaging}%
  \BibitemOpen
  \bibfield  {author} {\bibinfo {author} {\bibfnamefont {J.}~\bibnamefont
  {Itatani}}, \bibinfo {author} {\bibfnamefont {J.}~\bibnamefont {Levesque}},
  \bibinfo {author} {\bibfnamefont {D.}~\bibnamefont {Zeidler}}, \bibinfo
  {author} {\bibfnamefont {H.}~\bibnamefont {Niikura}}, \bibinfo {author}
  {\bibfnamefont {H.}~\bibnamefont {P\'{e}pin}}, \bibinfo {author}
  {\bibfnamefont {J.~C.}\ \bibnamefont {Kieffer}}, \bibinfo {author}
  {\bibfnamefont {P.~B.}\ \bibnamefont {Corkum}}, \ and\ \bibinfo {author}
  {\bibfnamefont {D.~M.}\ \bibnamefont {Villeneuve}},\ }\href@noop {}
  {\bibfield  {journal} {\bibinfo  {journal} {Nature}\ }\textbf {\bibinfo
  {volume} {432}},\ \bibinfo {pages} {867} (\bibinfo {year}
  {2004})}\BibitemShut {NoStop}%
\bibitem [{\citenamefont {Resch}\ \emph {et~al.}(2005)\citenamefont {Resch},
  \citenamefont {Walther},\ and\ \citenamefont
  {Zeilinger}}]{Resch_PRL05:ThreePhotonStateTomography}%
  \BibitemOpen
  \bibfield  {author} {\bibinfo {author} {\bibfnamefont {K.~J.}\ \bibnamefont
  {Resch}}, \bibinfo {author} {\bibfnamefont {P.}~\bibnamefont {Walther}}, \
  and\ \bibinfo {author} {\bibfnamefont {A.}~\bibnamefont {Zeilinger}},\ }\href
  {\doibase 10.1103/PhysRevLett.94.070402} {\bibfield  {journal} {\bibinfo
  {journal} {Phys. Rev. Lett.}\ }\textbf {\bibinfo {volume} {94}},\ \bibinfo
  {pages} {070402} (\bibinfo {year} {2005})}\BibitemShut {NoStop}%
\bibitem [{\citenamefont {S\"{o}derholm}\ \emph {et~al.}(2012)\citenamefont
  {S\"{o}derholm}, \citenamefont {Bj\"{o}rk}, \citenamefont {Klimov},
  \citenamefont {S\'{a}nchez-Soto},\ and\ \citenamefont
  {Leuchs}}]{Soderholm_NJP12:PolaTomography}%
  \BibitemOpen
  \bibfield  {author} {\bibinfo {author} {\bibfnamefont {J.}~\bibnamefont
  {S\"{o}derholm}}, \bibinfo {author} {\bibfnamefont {G.}~\bibnamefont
  {Bj\"{o}rk}}, \bibinfo {author} {\bibfnamefont {A.~B.}\ \bibnamefont
  {Klimov}}, \bibinfo {author} {\bibfnamefont {L.~L.}\ \bibnamefont
  {S\'{a}nchez-Soto}}, \ and\ \bibinfo {author} {\bibfnamefont
  {G.}~\bibnamefont {Leuchs}},\ }\href
  {http://stacks.iop.org/1367-2630/14/i=11/a=115014} {\bibfield  {journal}
  {\bibinfo  {journal} {New Journal of Physics}\ }\textbf {\bibinfo {volume}
  {14}},\ \bibinfo {pages} {115014} (\bibinfo {year} {2012})}\BibitemShut
  {NoStop}%
\bibitem [{\citenamefont {Sych}\ \emph {et~al.}(2012)\citenamefont {Sych},
  \citenamefont {\ifmmode \check{R}\else \v{R}\fi{}eh\'a\ifmmode~\check{c}\else
  \v{c}\fi{}ek}, \citenamefont {Hradil}, \citenamefont {Leuchs},\ and\
  \citenamefont {S\'anchez-Soto}}]{Sych_PRA12:InfoCompleteContVariable}%
  \BibitemOpen
  \bibfield  {author} {\bibinfo {author} {\bibfnamefont {D.}~\bibnamefont
  {Sych}}, \bibinfo {author} {\bibfnamefont {J.}~\bibnamefont {\ifmmode
  \check{R}\else \v{R}\fi{}eh\'a\ifmmode~\check{c}\else \v{c}\fi{}ek}},
  \bibinfo {author} {\bibfnamefont {Z.}~\bibnamefont {Hradil}}, \bibinfo
  {author} {\bibfnamefont {G.}~\bibnamefont {Leuchs}}, \ and\ \bibinfo {author}
  {\bibfnamefont {L.~L.}\ \bibnamefont {S\'anchez-Soto}},\ }\href {\doibase
  10.1103/PhysRevA.86.052123} {\bibfield  {journal} {\bibinfo  {journal} {Phys.
  Rev. A}\ }\textbf {\bibinfo {volume} {86}},\ \bibinfo {pages} {052123}
  (\bibinfo {year} {2012})}\BibitemShut {NoStop}%
\bibitem [{\citenamefont {Beck}(2000)}]{Beck_PRL00:QuanTomoDetArray}%
  \BibitemOpen
  \bibfield  {author} {\bibinfo {author} {\bibfnamefont {M.}~\bibnamefont
  {Beck}},\ }\href {\doibase 10.1103/PhysRevLett.84.5748} {\bibfield  {journal}
  {\bibinfo  {journal} {Phys. Rev. Lett.}\ }\textbf {\bibinfo {volume} {84}},\
  \bibinfo {pages} {5748} (\bibinfo {year} {2000})}\BibitemShut {NoStop}%
\bibitem [{\citenamefont {Beck}\ \emph {et~al.}(2001)\citenamefont {Beck},
  \citenamefont {Dorrer},\ and\ \citenamefont
  {Walmsley}}]{Beck_PRL01:QuanMeasureDetArray}%
  \BibitemOpen
  \bibfield  {author} {\bibinfo {author} {\bibfnamefont {M.}~\bibnamefont
  {Beck}}, \bibinfo {author} {\bibfnamefont {C.}~\bibnamefont {Dorrer}}, \ and\
  \bibinfo {author} {\bibfnamefont {I.~A.}\ \bibnamefont {Walmsley}},\ }\href
  {\doibase 10.1103/PhysRevLett.87.253601} {\bibfield  {journal} {\bibinfo
  {journal} {Phys. Rev. Lett.}\ }\textbf {\bibinfo {volume} {87}},\ \bibinfo
  {pages} {253601} (\bibinfo {year} {2001})}\BibitemShut {NoStop}%
\bibitem [{\citenamefont {Dawes}\ \emph {et~al.}(2003)\citenamefont {Dawes},
  \citenamefont {Beck},\ and\ \citenamefont
  {Banaszek}}]{Dawes_PRA03:quantumTomowDetArray}%
  \BibitemOpen
  \bibfield  {author} {\bibinfo {author} {\bibfnamefont {A.~M.}\ \bibnamefont
  {Dawes}}, \bibinfo {author} {\bibfnamefont {M.}~\bibnamefont {Beck}}, \ and\
  \bibinfo {author} {\bibfnamefont {K.}~\bibnamefont {Banaszek}},\ }\href
  {\doibase 10.1103/PhysRevA.67.032102} {\bibfield  {journal} {\bibinfo
  {journal} {Phys. Rev. A}\ }\textbf {\bibinfo {volume} {67}},\ \bibinfo
  {pages} {032102} (\bibinfo {year} {2003})}\BibitemShut {NoStop}%
\bibitem [{\citenamefont {Lundeen}\ and\ \citenamefont
  {Bamber}(2012)}]{Lundeen_PRL12:WeakMeasureGeneral}%
  \BibitemOpen
  \bibfield  {author} {\bibinfo {author} {\bibfnamefont {J.~S.}\ \bibnamefont
  {Lundeen}}\ and\ \bibinfo {author} {\bibfnamefont {C.}~\bibnamefont
  {Bamber}},\ }\href {\doibase 10.1103/PhysRevLett.108.070402} {\bibfield
  {journal} {\bibinfo  {journal} {Phys. Rev. Lett.}\ }\textbf {\bibinfo
  {volume} {108}},\ \bibinfo {pages} {070402} (\bibinfo {year}
  {2012})}\BibitemShut {NoStop}%
\bibitem [{\citenamefont {Shengjun}(2013)}]{Wu_SciRep:StateTomography}%
  \BibitemOpen
  \bibfield  {author} {\bibinfo {author} {\bibfnamefont {W.}~\bibnamefont
  {Shengjun}},\ }\href {\doibase 10.1038/srep01193} {\bibfield  {journal}
  {\bibinfo  {journal} {Scientific Reports}\ }\textbf {\bibinfo {volume} {3}},\
  \bibinfo {pages} {1193} (\bibinfo {year} {2013})}\BibitemShut {NoStop}%
\bibitem [{\citenamefont {Salvail}\ \emph {et~al.}(2013)\citenamefont
  {Salvail}, \citenamefont {Agnew}, \citenamefont {Johnson}, \citenamefont
  {Bolduc}, \citenamefont {Leach},\ and\ \citenamefont
  {Boyd}}]{Saivail_NatPhon13:PolarizationDirectMeasure}%
  \BibitemOpen
  \bibfield  {author} {\bibinfo {author} {\bibfnamefont {J.~Z.}\ \bibnamefont
  {Salvail}}, \bibinfo {author} {\bibfnamefont {M.}~\bibnamefont {Agnew}},
  \bibinfo {author} {\bibfnamefont {A.~S.}\ \bibnamefont {Johnson}}, \bibinfo
  {author} {\bibfnamefont {E.}~\bibnamefont {Bolduc}}, \bibinfo {author}
  {\bibfnamefont {J.}~\bibnamefont {Leach}}, \ and\ \bibinfo {author}
  {\bibfnamefont {R.~W.}\ \bibnamefont {Boyd}},\ }\href@noop {} {\bibfield
  {journal} {\bibinfo  {journal} {Nature Photonics}\ }\textbf {\bibinfo
  {volume} {7}},\ \bibinfo {pages} {316} (\bibinfo {year} {2013})}\BibitemShut
  {NoStop}%
\bibitem [{\citenamefont {Malik}\ \emph {et~al.}(2014)\citenamefont {Malik},
  \citenamefont {Mirhosseini}, \citenamefont {Lavery}, \citenamefont {Leach},
  \citenamefont {Padgett},\ and\ \citenamefont
  {Boyd}}]{Malik_13tj:DirectMeasureOAM}%
  \BibitemOpen
  \bibfield  {author} {\bibinfo {author} {\bibfnamefont {M.}~\bibnamefont
  {Malik}}, \bibinfo {author} {\bibfnamefont {M.}~\bibnamefont {Mirhosseini}},
  \bibinfo {author} {\bibfnamefont {M.~P.~J.}\ \bibnamefont {Lavery}}, \bibinfo
  {author} {\bibfnamefont {J.}~\bibnamefont {Leach}}, \bibinfo {author}
  {\bibfnamefont {M.~J.}\ \bibnamefont {Padgett}}, \ and\ \bibinfo {author}
  {\bibfnamefont {R.~W.}\ \bibnamefont {Boyd}},\ }\href@noop {} {\bibfield
  {journal} {\bibinfo  {journal} {Nature Communications}\ }\textbf {\bibinfo
  {volume} {4}},\ \bibinfo {pages} {3115} (\bibinfo {year} {2014})}\BibitemShut
  {NoStop}%
\bibitem [{\citenamefont {Aharonov}\ \emph {et~al.}(1988)\citenamefont
  {Aharonov}, \citenamefont {Albert},\ and\ \citenamefont
  {Vaidman}}]{Aharonov_PRL88:WeakMeasurementSpin}%
  \BibitemOpen
  \bibfield  {author} {\bibinfo {author} {\bibfnamefont {Y.}~\bibnamefont
  {Aharonov}}, \bibinfo {author} {\bibfnamefont {D.~Z.}\ \bibnamefont
  {Albert}}, \ and\ \bibinfo {author} {\bibfnamefont {L.}~\bibnamefont
  {Vaidman}},\ }\href {\doibase 10.1103/PhysRevLett.60.1351} {\bibfield
  {journal} {\bibinfo  {journal} {Phys. Rev. Lett.}\ }\textbf {\bibinfo
  {volume} {60}},\ \bibinfo {pages} {1351} (\bibinfo {year}
  {1988})}\BibitemShut {NoStop}%
\bibitem [{\citenamefont {Duck}\ \emph {et~al.}(1989)\citenamefont {Duck},
  \citenamefont {Stevenson},\ and\ \citenamefont
  {Sudarshan}}]{Duck_PRD89:WeakMeasurement}%
  \BibitemOpen
  \bibfield  {author} {\bibinfo {author} {\bibfnamefont {I.~M.}\ \bibnamefont
  {Duck}}, \bibinfo {author} {\bibfnamefont {P.~M.}\ \bibnamefont {Stevenson}},
  \ and\ \bibinfo {author} {\bibfnamefont {E.~C.~G.}\ \bibnamefont
  {Sudarshan}},\ }\href {\doibase 10.1103/PhysRevD.40.2112} {\bibfield
  {journal} {\bibinfo  {journal} {Phys. Rev. D}\ }\textbf {\bibinfo {volume}
  {40}},\ \bibinfo {pages} {2112} (\bibinfo {year} {1989})}\BibitemShut
  {NoStop}%
\bibitem [{\citenamefont {Ritchie}\ \emph {et~al.}(1991)\citenamefont
  {Ritchie}, \citenamefont {Story},\ and\ \citenamefont
  {Hulet}}]{Ritchie_PRL91:RealizationWeakMeasurement}%
  \BibitemOpen
  \bibfield  {author} {\bibinfo {author} {\bibfnamefont {N.~W.~M.}\
  \bibnamefont {Ritchie}}, \bibinfo {author} {\bibfnamefont {J.~G.}\
  \bibnamefont {Story}}, \ and\ \bibinfo {author} {\bibfnamefont {R.~G.}\
  \bibnamefont {Hulet}},\ }\href {\doibase 10.1103/PhysRevLett.66.1107}
  {\bibfield  {journal} {\bibinfo  {journal} {Phys. Rev. Lett.}\ }\textbf
  {\bibinfo {volume} {66}},\ \bibinfo {pages} {1107} (\bibinfo {year}
  {1991})}\BibitemShut {NoStop}%
\bibitem [{\citenamefont
  {Johansen}(2004)}]{Johansen_PRL04:WeakMeasurementArbitraryProbeStates}%
  \BibitemOpen
  \bibfield  {author} {\bibinfo {author} {\bibfnamefont {L.~M.}\ \bibnamefont
  {Johansen}},\ }\href {\doibase 10.1103/PhysRevLett.93.120402} {\bibfield
  {journal} {\bibinfo  {journal} {Phys. Rev. Lett.}\ }\textbf {\bibinfo
  {volume} {93}},\ \bibinfo {pages} {120402} (\bibinfo {year}
  {2004})}\BibitemShut {NoStop}%
\bibitem [{\citenamefont {Hosten}\ and\ \citenamefont
  {Kwiat}(2004)}]{Hosten_Science04:SpinHallEff_weakMeasure}%
  \BibitemOpen
  \bibfield  {author} {\bibinfo {author} {\bibfnamefont {O.}~\bibnamefont
  {Hosten}}\ and\ \bibinfo {author} {\bibfnamefont {P.}~\bibnamefont {Kwiat}},\
  }\href@noop {} {\bibfield  {journal} {\bibinfo  {journal} {Science}\ }\textbf
  {\bibinfo {volume} {319}},\ \bibinfo {pages} {787} (\bibinfo {year}
  {2004})}\BibitemShut {NoStop}%
\bibitem [{\citenamefont {Solli}\ \emph {et~al.}(2004)\citenamefont {Solli},
  \citenamefont {McCormick}, \citenamefont {Chiao}, \citenamefont {Popescu},\
  and\ \citenamefont {Hickmann}}]{Solli_PRL04:FLSL_GeneralizedWeakValue}%
  \BibitemOpen
  \bibfield  {author} {\bibinfo {author} {\bibfnamefont {D.~R.}\ \bibnamefont
  {Solli}}, \bibinfo {author} {\bibfnamefont {C.~F.}\ \bibnamefont
  {McCormick}}, \bibinfo {author} {\bibfnamefont {R.~Y.}\ \bibnamefont
  {Chiao}}, \bibinfo {author} {\bibfnamefont {S.}~\bibnamefont {Popescu}}, \
  and\ \bibinfo {author} {\bibfnamefont {J.~M.}\ \bibnamefont {Hickmann}},\
  }\href {\doibase 10.1103/PhysRevLett.92.043601} {\bibfield  {journal}
  {\bibinfo  {journal} {Phys. Rev. Lett.}\ }\textbf {\bibinfo {volume} {92}},\
  \bibinfo {pages} {043601} (\bibinfo {year} {2004})}\BibitemShut {NoStop}%
\bibitem [{\citenamefont {Dixon}\ \emph {et~al.}(2009)\citenamefont {Dixon},
  \citenamefont {Starling}, \citenamefont {Jordan},\ and\ \citenamefont
  {Howell}}]{DixonPRL09:WeakDeflection}%
  \BibitemOpen
  \bibfield  {author} {\bibinfo {author} {\bibfnamefont {P.~B.}\ \bibnamefont
  {Dixon}}, \bibinfo {author} {\bibfnamefont {D.~J.}\ \bibnamefont {Starling}},
  \bibinfo {author} {\bibfnamefont {A.~N.}\ \bibnamefont {Jordan}}, \ and\
  \bibinfo {author} {\bibfnamefont {J.~C.}\ \bibnamefont {Howell}},\ }\href
  {\doibase 10.1103/PhysRevLett.102.173601} {\bibfield  {journal} {\bibinfo
  {journal} {Phys. Rev. Lett.}\ }\textbf {\bibinfo {volume} {102}},\ \bibinfo
  {pages} {173601} (\bibinfo {year} {2009})}\BibitemShut {NoStop}%
\bibitem [{\citenamefont {Feizpour}\ \emph {et~al.}(2011)\citenamefont
  {Feizpour}, \citenamefont {Xing},\ and\ \citenamefont
  {Steinberg}}]{Feizpour_PRL11:WeakSinglePhotonNonlinearity}%
  \BibitemOpen
  \bibfield  {author} {\bibinfo {author} {\bibfnamefont {A.}~\bibnamefont
  {Feizpour}}, \bibinfo {author} {\bibfnamefont {X.}~\bibnamefont {Xing}}, \
  and\ \bibinfo {author} {\bibfnamefont {A.~M.}\ \bibnamefont {Steinberg}},\
  }\href {\doibase 10.1103/PhysRevLett.107.133603} {\bibfield  {journal}
  {\bibinfo  {journal} {Phys. Rev. Lett.}\ }\textbf {\bibinfo {volume} {107}},\
  \bibinfo {pages} {133603} (\bibinfo {year} {2011})}\BibitemShut {NoStop}%
\bibitem [{\citenamefont {Kocsis}\ \emph {et~al.}(2011)\citenamefont {Kocsis},
  \citenamefont {Braverman}, \citenamefont {Ravets}, \citenamefont {Stevens},
  \citenamefont {Mirin}, \citenamefont {Shalm},\ and\ \citenamefont
  {Steinberg}}]{Kocsis_Science12:EMCCD_entanglement}%
  \BibitemOpen
  \bibfield  {author} {\bibinfo {author} {\bibfnamefont {S.}~\bibnamefont
  {Kocsis}}, \bibinfo {author} {\bibfnamefont {B.}~\bibnamefont {Braverman}},
  \bibinfo {author} {\bibfnamefont {S.}~\bibnamefont {Ravets}}, \bibinfo
  {author} {\bibfnamefont {M.~J.}\ \bibnamefont {Stevens}}, \bibinfo {author}
  {\bibfnamefont {R.~P.}\ \bibnamefont {Mirin}}, \bibinfo {author}
  {\bibfnamefont {L.~K.}\ \bibnamefont {Shalm}}, \ and\ \bibinfo {author}
  {\bibfnamefont {A.~M.}\ \bibnamefont {Steinberg}},\ }\href {\doibase
  10.1126/science.1202218} {\bibfield  {journal} {\bibinfo  {journal}
  {Science}\ }\textbf {\bibinfo {volume} {332}},\ \bibinfo {pages} {1170}
  (\bibinfo {year} {2011})}\BibitemShut {NoStop}%
\bibitem [{\citenamefont {Dressel}\ \emph {et~al.}(2014)\citenamefont
  {Dressel}, \citenamefont {Malik}, \citenamefont {Miatto}, \citenamefont
  {Jordan},\ and\ \citenamefont {Boyd}}]{Dressel_13ti:WeakReview}%
  \BibitemOpen
  \bibfield  {author} {\bibinfo {author} {\bibfnamefont {J.}~\bibnamefont
  {Dressel}}, \bibinfo {author} {\bibfnamefont {M.}~\bibnamefont {Malik}},
  \bibinfo {author} {\bibfnamefont {F.~M.}\ \bibnamefont {Miatto}}, \bibinfo
  {author} {\bibfnamefont {A.~N.}\ \bibnamefont {Jordan}}, \ and\ \bibinfo
  {author} {\bibfnamefont {R.~W.}\ \bibnamefont {Boyd}},\ }\href@noop {}
  {\bibfield  {journal} {\bibinfo  {journal} {Rev. Mod. Phys.}\ }\textbf
  {\bibinfo {volume} {86}},\ \bibinfo {pages} {307} (\bibinfo {year}
  {2014})}\BibitemShut {NoStop}%
\bibitem [{\citenamefont {Davis}\ \emph {et~al.}(1999)\citenamefont {Davis},
  \citenamefont {Cottrell}, \citenamefont {Campos}, \citenamefont {Yzuel},\
  and\ \citenamefont {Moreno}}]{Davis_AO99:AmpModwphaseSLM}%
  \BibitemOpen
  \bibfield  {author} {\bibinfo {author} {\bibfnamefont {J.~A.}\ \bibnamefont
  {Davis}}, \bibinfo {author} {\bibfnamefont {D.~M.}\ \bibnamefont {Cottrell}},
  \bibinfo {author} {\bibfnamefont {J.}~\bibnamefont {Campos}}, \bibinfo
  {author} {\bibfnamefont {M.~J.}\ \bibnamefont {Yzuel}}, \ and\ \bibinfo
  {author} {\bibfnamefont {I.}~\bibnamefont {Moreno}},\ }\href {\doibase
  10.1364/AO.38.005004} {\bibfield  {journal} {\bibinfo  {journal} {Appl.
  Opt.}\ }\textbf {\bibinfo {volume} {38}},\ \bibinfo {pages} {5004} (\bibinfo
  {year} {1999})}\BibitemShut {NoStop}%
\bibitem [{\citenamefont {Arriz\'{o}n}\ \emph {et~al.}(2007)\citenamefont
  {Arriz\'{o}n}, \citenamefont {Ruiz}, \citenamefont {Carrada},\ and\
  \citenamefont {Gonz\'{a}lez}}]{Arrizon_JOSAA07:SLMgenerateComplexField}%
  \BibitemOpen
  \bibfield  {author} {\bibinfo {author} {\bibfnamefont {V.}~\bibnamefont
  {Arriz\'{o}n}}, \bibinfo {author} {\bibfnamefont {U.}~\bibnamefont {Ruiz}},
  \bibinfo {author} {\bibfnamefont {R.}~\bibnamefont {Carrada}}, \ and\
  \bibinfo {author} {\bibfnamefont {L.~A.}\ \bibnamefont {Gonz\'{a}lez}},\
  }\href {\doibase 10.1364/JOSAA.24.003500} {\bibfield  {journal} {\bibinfo
  {journal} {J. Opt. Soc. Am. A}\ }\textbf {\bibinfo {volume} {24}},\ \bibinfo
  {pages} {3500} (\bibinfo {year} {2007})}\BibitemShut {NoStop}%
\bibitem [{\citenamefont {Smartt}\ and\ \citenamefont
  {Strong}(1972)}]{Smartt_JOSA72:PDI}%
  \BibitemOpen
  \bibfield  {author} {\bibinfo {author} {\bibfnamefont {R.~N.}\ \bibnamefont
  {Smartt}}\ and\ \bibinfo {author} {\bibfnamefont {J.}~\bibnamefont
  {Strong}},\ }\href@noop {} {\bibfield  {journal} {\bibinfo  {journal} {J.
  Opt. Soc. Am.}\ }\textbf {\bibinfo {volume} {62}},\ \bibinfo {pages} {737}
  (\bibinfo {year} {1972})}\BibitemShut {NoStop}%
\bibitem [{\citenamefont {Yao}\ and\ \citenamefont
  {Padgett}(2011)}]{Yao_AOP11:OAMreview}%
  \BibitemOpen
  \bibfield  {author} {\bibinfo {author} {\bibfnamefont {A.~M.}\ \bibnamefont
  {Yao}}\ and\ \bibinfo {author} {\bibfnamefont {M.~J.}\ \bibnamefont
  {Padgett}},\ }\href {\doibase 10.1364/AOP.3.000161} {\bibfield  {journal}
  {\bibinfo  {journal} {Adv. Opt. Photon.}\ }\textbf {\bibinfo {volume} {3}},\
  \bibinfo {pages} {161} (\bibinfo {year} {2011})}\BibitemShut {NoStop}%
\bibitem [{\citenamefont {Leach}\ \emph {et~al.}(2010)\citenamefont {Leach},
  \citenamefont {Jack}, \citenamefont {Romero}, \citenamefont {Jha},
  \citenamefont {Yao}, \citenamefont {Franke-Arnold}, \citenamefont {Ireland},
  \citenamefont {Boyd}, \citenamefont {Barnett},\ and\ \citenamefont
  {Padgett}}]{Leach_Science10:QuantumCorrelationOAM}%
  \BibitemOpen
  \bibfield  {author} {\bibinfo {author} {\bibfnamefont {J.}~\bibnamefont
  {Leach}}, \bibinfo {author} {\bibfnamefont {B.}~\bibnamefont {Jack}},
  \bibinfo {author} {\bibfnamefont {J.}~\bibnamefont {Romero}}, \bibinfo
  {author} {\bibfnamefont {A.~K.}\ \bibnamefont {Jha}}, \bibinfo {author}
  {\bibfnamefont {A.~M.}\ \bibnamefont {Yao}}, \bibinfo {author} {\bibfnamefont
  {S.}~\bibnamefont {Franke-Arnold}}, \bibinfo {author} {\bibfnamefont {D.~G.}\
  \bibnamefont {Ireland}}, \bibinfo {author} {\bibfnamefont {R.~W.}\
  \bibnamefont {Boyd}}, \bibinfo {author} {\bibfnamefont {S.~M.}\ \bibnamefont
  {Barnett}}, \ and\ \bibinfo {author} {\bibfnamefont {M.~J.}\ \bibnamefont
  {Padgett}},\ }\href@noop {} {\bibfield  {journal} {\bibinfo  {journal}
  {Science}\ }\textbf {\bibinfo {volume} {329}},\ \bibinfo {pages} {662}
  (\bibinfo {year} {2010})}\BibitemShut {NoStop}%
\bibitem [{\citenamefont {Dada}\ \emph {et~al.}(2011)\citenamefont {Dada},
  \citenamefont {Leach}, \citenamefont {Buller}, \citenamefont {Padgett},\ and\
  \citenamefont {Andersson}}]{Dada_NatPhys11:HighDimEntanglement}%
  \BibitemOpen
  \bibfield  {author} {\bibinfo {author} {\bibfnamefont {A.~C.}\ \bibnamefont
  {Dada}}, \bibinfo {author} {\bibfnamefont {J.}~\bibnamefont {Leach}},
  \bibinfo {author} {\bibfnamefont {G.~S.}\ \bibnamefont {Buller}}, \bibinfo
  {author} {\bibfnamefont {M.~J.}\ \bibnamefont {Padgett}}, \ and\ \bibinfo
  {author} {\bibfnamefont {E.}~\bibnamefont {Andersson}},\ }\href@noop {}
  {\bibfield  {journal} {\bibinfo  {journal} {Nature Physics}\ }\textbf
  {\bibinfo {volume} {7}},\ \bibinfo {pages} {677} (\bibinfo {year}
  {2011})}\BibitemShut {NoStop}%
\bibitem [{\citenamefont {Malik}\ \emph {et~al.}(2012)\citenamefont {Malik},
  \citenamefont {O'Sullivan}, \citenamefont {Rodenburg}, \citenamefont
  {Mirhosseini}, \citenamefont {Leach}, \citenamefont {Lavery}, \citenamefont
  {Padgett},\ and\ \citenamefont {Boyd}}]{Malik_OE12:Turbulence}%
  \BibitemOpen
  \bibfield  {author} {\bibinfo {author} {\bibfnamefont {M.}~\bibnamefont
  {Malik}}, \bibinfo {author} {\bibfnamefont {M.~N.}\ \bibnamefont
  {O'Sullivan}}, \bibinfo {author} {\bibfnamefont {B.}~\bibnamefont
  {Rodenburg}}, \bibinfo {author} {\bibfnamefont {M.}~\bibnamefont
  {Mirhosseini}}, \bibinfo {author} {\bibfnamefont {J.}~\bibnamefont {Leach}},
  \bibinfo {author} {\bibfnamefont {M.~P.~J.}\ \bibnamefont {Lavery}}, \bibinfo
  {author} {\bibfnamefont {M.~J.}\ \bibnamefont {Padgett}}, \ and\ \bibinfo
  {author} {\bibfnamefont {R.~W.}\ \bibnamefont {Boyd}},\ }\href@noop {}
  {\bibfield  {journal} {\bibinfo  {journal} {Optics Express}\ }\textbf
  {\bibinfo {volume} {20}},\ \bibinfo {pages} {13195} (\bibinfo {year}
  {2012})}\BibitemShut {NoStop}%
\bibitem [{\citenamefont {Fickler}\ \emph {et~al.}(2013)\citenamefont
  {Fickler}, \citenamefont {Krenn}, \citenamefont {Lapkiewicz}, \citenamefont
  {Ramelow},\ and\ \citenamefont
  {Zeilinger}}]{Fickler_SciRep:RealTimeImagingEntanglement}%
  \BibitemOpen
  \bibfield  {author} {\bibinfo {author} {\bibfnamefont {R.}~\bibnamefont
  {Fickler}}, \bibinfo {author} {\bibfnamefont {M.}~\bibnamefont {Krenn}},
  \bibinfo {author} {\bibfnamefont {R.}~\bibnamefont {Lapkiewicz}}, \bibinfo
  {author} {\bibfnamefont {S.}~\bibnamefont {Ramelow}}, \ and\ \bibinfo
  {author} {\bibfnamefont {A.}~\bibnamefont {Zeilinger}},\ }\href {\doibase
  10.1038/srep01914} {\bibfield  {journal} {\bibinfo  {journal} {Scientific
  Reports}\ }\textbf {\bibinfo {volume} {3}},\ \bibinfo {pages} {1914}
  (\bibinfo {year} {2013})}\BibitemShut {NoStop}%
\bibitem [{\citenamefont {Edgar}\ \emph {et~al.}(2012)\citenamefont {Edgar},
  \citenamefont {Tasca}, \citenamefont {Izdebski}, \citenamefont {Warburton},
  \citenamefont {Leach}, \citenamefont {Agnew}, \citenamefont {Buller},
  \citenamefont {Boyd},\ and\ \citenamefont
  {Padgett}}]{Edgar_NatComm12:EMCCD_entanglement}%
  \BibitemOpen
  \bibfield  {author} {\bibinfo {author} {\bibfnamefont {M.}~\bibnamefont
  {Edgar}}, \bibinfo {author} {\bibfnamefont {D.}~\bibnamefont {Tasca}},
  \bibinfo {author} {\bibfnamefont {F.}~\bibnamefont {Izdebski}}, \bibinfo
  {author} {\bibfnamefont {R.}~\bibnamefont {Warburton}}, \bibinfo {author}
  {\bibfnamefont {J.}~\bibnamefont {Leach}}, \bibinfo {author} {\bibfnamefont
  {M.}~\bibnamefont {Agnew}}, \bibinfo {author} {\bibfnamefont
  {G.}~\bibnamefont {Buller}}, \bibinfo {author} {\bibfnamefont
  {R.}~\bibnamefont {Boyd}}, \ and\ \bibinfo {author} {\bibfnamefont
  {M.}~\bibnamefont {Padgett}},\ }\href {\doibase 10.1038/ncomms1988}
  {\bibfield  {journal} {\bibinfo  {journal} {Nature Commun.}\ }\textbf
  {\bibinfo {volume} {3}},\ \bibinfo {pages} {984} (\bibinfo {year}
  {2012})}\BibitemShut {NoStop}%
\end{thebibliography}
%merlin.mbs apsrev4-1.bst 2010-07-25 4.21a (PWD, AO, DPC) hacked
%Control: key (0)
%Control: author (72) initials jnrlst
%Control: editor formatted (1) identically to author
%Control: production of article title (-1) disabled
%Control: page (0) single
%Control: year (1) truncated
%Control: production of eprint (0) enabled
%

\end{document}